\input epsf
\documentstyle[aps,12pt]{revtex}
\begin{document}
\author{M. V. Voitikova}
\title{STRANGE ATTRACTORS IN IMMUNOLOGY OF TUMOR GROWTH.}
\address{Institute of Molecular\&Atomic Physics, Academy of sciences of Belarus,\\
F.Skaryna av. 70, Minsk 220072, Belarus}
\maketitle

\abstract{The time delayed cytotoxic T-lymphocyte response on the tumor
growth has 
been developed on the basis of discrete approximation (2-dimensional map). The 
growth kinetic has been described by logistic law with growth rate being the 
bifurcation parameter. Increase in the growth rate results in instability of the 
tumor state and causes period-doubling bifurcations in the immune+tumor system. 
For larger values of tumor growth rate a strange attractor has been observed. 
The model proposed is able to describe the metastable-state production when time 
series data of the immune state and the number of tumor cells are irregular and 
unpredictable. This metastatic disease may be caused not by exterior (medical) 
factors, but interior density dependent ones.}

\section{Introduction}

An immune system as an evolution formed one has a very high degree of
complexity and demonstrates the combination of chaotic and deterministic
properties. Different host defense cells are able to suppress the growth
tumor cells or destroy them. However, in some experimental and clinical
cases it has been observed that stimulation of an immune system by
immunotherapy results in stimulation of a tumor cell growth but not
suppression. Dormancy, regression, recurrence are possible stages of tumor
growth [1] which represent static state, reduction and expansion of the
tumor respectively. By dormancy we mean that a tumor, though being variable,
does not grow, i.e. the volume or cell number does not change with time. It
is frequently presumed that tumor cells do not grow because immune effector
cells kill tumor cells at a rate equal to that at which they are generated.
Regression requires that the tumor shrinks, and recurrence means that the
dormant tumor is now forced into a new growth phase. In some clinical
experiments a tumor volume follows an irregular cycle, whereas at other
cases it fluctuates drastically [2]. The discovery that fluctuating in time
patterns in immunological systems may be described by deterministic chaos
has revived the tumor growth study. A tumor also is able to generate new
clones resistant to cytolytic mechanisms, or is able to express the receptor
to growth factor, or penetrate into another tissues. So, a tumor is a rather
complicated system that can change its properties. Order and chaos in
immunology of a tumor growth can be linked by hierarchical organization.
Chaotic systems are entirely deterministic without random inputs in spite of
showing nonperiodic and noise-like motion. Furthermore, they exhibit
sensitive dependence on initial conditions, i.e. nearby trajectories
separate exponentially. Chaos in biological systems was identified and
modeled in several kinds of situations, including ecology, tumor growth, and
neural systems. The nonlinear character of biological populations and their
fractal properties are now recognized as properties of nature. In this
context, biophysical investigation of nonlinear systems is able to explain
paradoxical phenomena in oncology and predict possible scenarios of a
disease evolution and treatment.

In this paper, we have both investigated the immune response on the tumor
growth on the base of 2 - dimensional maps and  demonstrated the possibility
of metastable states of an immune+tumor system. 2 - dimensional discrete
maps are very useful for modeling realistic systems when there is no enough
empirical data to make a reasonable approximation and/or data is very noisy.
Population biologists often model dynamical population systems using
discrete-time model, which yields a prediction of a population at the next
cycle based on the population at the previous cycle.

After visualization and bifurcation analysis of the model it will be shown
that for a biological reasonable region of  parameters the  immune response
delay leads to a metastable state which in mathematical description
represent the strange attractor in the immunology of tumor growth. We have
been dealt with first order difference equation systems with discretely
changing variables.

\section{Description of the immunological model.}

We define the immune competence as the elimination capacity of an immune
system with respect to tumor cells. The tumor, which cells are immunogenetic
and attacked by cytotoxic effector cells, can trigger processes in the
immune system leading to competence against tumor cells. Activated
nonspecific precursor immune cells generate specific effector cells, e.g.
natural killers or cytotoxic T-lymphocyte (CTL) playing a dominant role.
Applying the Euler's method to the continuous-time system [1] yields the
discrete-time model. There is time delay $\Delta \tau $ between the moment
of stimulation of nonspecific precursor immune cells and generation of
specific effector cells. Furthermore, for the fast growing tumor cell
division time can be about few days. The immune+tumor system is composed of
two distinct populations,  $x$ and $y$. Now we approximate $x_n$ and $y_n$
as constants in the time interval from  $t_n=n\Delta \tau $ to  $%
t_{n+1}=(n+1)\Delta \tau $ . In numerous studies it was found that the
growth of tumor cell population is exponential for small quantities of tumor
cells, but growth is slower at large population size. The interaction
between effector cell concentration $x_n$ and tumor cell concentration $y_n$
can be described by the difference equations:

\begin{eqnarray}
x_{n+1} &=&x_n+\left( \frac{cx_ny_n}{\varphi +y_n}-\beta x_ny_n-\gamma
x_n+j\right) \Delta \tau =F(x_n,y_n)~,\quad  \label{1} \\
y_{n+1} &=&y_n+\left( \alpha y_n-\theta y_n^2-ax_ny_n\right) \Delta \tau
=G(x_n,y_n)~,\quad  \nonumber
\end{eqnarray}

here $\alpha $ is the logistic growth rate of tumor population, $\beta $ and 
$a$ are the rate of effector and tumor cells inactivation respectively, $%
\theta $ is the parameter of competition for resources (glucose, oxygen,
etc.), $c$ is the rate at which cytotoxic effector cells accumulates in the
region of tumor cells, and $\Delta \tau $ - takes into account times
necessary for tumor cell division and molecule production, proliferation,
differentiation of immune cells, transport, ets.

Analytically the system (1) in concentration scales has 1 to 4 non-trivial
steady states. One of them is the excluding point $A$ with coordinates $%
x_A=j/\gamma ,y_a=0$.  The point $C$ separates the basins of attraction for $%
B$ and $D$ attractors, so that

\begin{eqnarray}
x^{*} &=&F(x^{*},y^{*}),  \label{2} \\
y^{*} &=&G(x^{*},y^{*}).  \nonumber
\end{eqnarray}

The stability properties at the fixed points we investigate by evaluation of
the eigenvalues of the community matrix

\begin{equation}
\Gamma =\left( 
\begin{array}{cc}
\partial F_x & \partial G_x \\ 
\partial F_y & \partial G_y
\end{array}
\right) =\left( 
\begin{array}{cc}
1-\gamma -\beta y^{*}+cy^{*}/(\phi +y^{*}) & -\beta x^{*}-cx^{*}y^{*}/(\phi
+y^{*})^2 \\ 
-ay^{*} & 1+\alpha -ax^{*}-2\theta y^{*}
\end{array}
\right)  \label{3}
\end{equation}

In addition, we would note that  equilibrium points (constant stable
solutions) of continues-time systems [1] correspond to  fixed points of a
discrete-time model. For the point $A$ the associated eigenvalue equation, $%
\left| \Gamma (A)-\lambda I\right| =0$ , has two solutions,  $\lambda
_1=1-\gamma $ and  $\lambda _2=1+\alpha -aj/\gamma .$ The attractor $A$ will
be stable if both $\left| \lambda _{1,2}\right| \prec 1$ (normal wound
healing). The stability of $A$ steady state depends upon relative values of
the parameters $\alpha $, $j$, and $\gamma $. For example, the increase in
the logistic rate $\alpha $ or $\gamma $ lead to instability of the $A$
steady state, and transcritical bifurcation involving $A$ and $B$ steady
states takes place. The stable steady state $B(x_b,y_b)$ (tumor dormancy) is
characterized by a relatively low tumor cells amount. The stable steady
state $D(x_d,y_d)$ is characterized by a relatively high tumor and low
effector-cell amounts and corresponds to an uncontrolled tumor growth. The
steady state $C(x_c,y_c)$ separates the basins of attraction for the $B$ and 
$D$ attractors (Fig.1). The coordinates of steady states $B$, $C$, and $D$
can be obtained by finding the positive roots of the equation:

\begin{equation}
a_3y^3+a_2^{}y^2+a_1y+a_0=0  \label{4}
\end{equation}
and 
\[
x=(\alpha -\theta y)/a, 
\]
where 
\[
a_0=\varphi (aj-\alpha \gamma ),a_1=\alpha (c-\beta \varphi -\gamma
)+aj+\varphi \gamma \theta ,a_2=\theta (\gamma +\beta \varphi -c)-\alpha
\theta ,a_3=\beta \theta ,\Delta \tau =1 
\]

The corresponding eigenvalues for steady states $B$, $C$, and $D$ are given
by $\lambda _{1,2}$ $=\sigma /2\pm \sqrt{\sigma ^2/4-\Delta }$ respectively,
where $\sigma $ $=\partial F_x$ $+\partial G_y$ , $\Delta =\partial
F_x\partial G_y-\partial F_y\partial G_x$. When parameters of the
immune+tumor system are varied so that one or two eigenvalues of steady
states $B$, $D$ cross the stability boundary $\left| \lambda _{1,2}\right| =1
$, characteristic bifurcations occur. It should be noted, that  $\left|
\lambda _{1,2}\right| $ $\succ 1$ for point $C$ everywhere. In this
situation two outcomes can be realized depending on initial conditions:

i) tumor dormancy - there is a high effector cell level and tumor presence
is reduced but not eliminated (the stable steady state $B$)

i i) uncontrolled tumor growth and immunological paralysis (the system
approaches the stable steady state $D$).

A phase portrait of the system is presented at the Fig. 1 for the parameters
estimated in [1] for BCL lymphoma in the spleen of chimeric mice : $\alpha
=1.3$ , $a=0.1$, $c=0.1$, $\beta =3\cdot 10^{-4},$ $\gamma =4\cdot 10^{-2}$, 
$\theta =20$, $\theta =2\cdot 10^{-3}$, and $j=10^{-2}$ . This portrait for
small parameter $\alpha $ is similar to ones for differential equations
without time delay [1]. It should be noted, that a tumor has a lot of
possibilities to overcome limitations of growth. Tumor cells change their
genetics, penetrate into other organs, become insensitive for attack of
effector cells, ets. This changing we associate with an increase in $\alpha $
, $\beta $ and decrease in $a$ and $c$. We can see that $\alpha $ is the
bifurcation parameter in controlling the stability of solutions, so there
are different stationary periodic and chaotic attractors depending on the
parameters values. For $\alpha <$1.0 the point $A$ is stable , and for $%
\alpha <$2.0 the points $B$ and $D$ are stable whereas the point $A$ is
unstable. As it is well known for the logistic map, the instability of $B$
and $D$ appears for $\alpha $ $>$2.2, in which a bifurcation takes place.
Chaotic dynamics occurs after a period-doubling-bifurcation scenario shown
in the Fig. 2. A dynamical system is called chaotic if at least one Lyapunov
exponent is positive and chaotic strange attractor covers a region with
fractal dimension. Using 100.000 time steps for parameter $\alpha =2.7$
(Fig.3) we got Lyapunov exponents: $\Lambda _1=0.35$, $\Lambda _2=-0.388$,
and dimension $D_L=1.92.$ It should be noted that the range of chaotic
dynamics depends on other relevant parameters.

\section{ Conclusions.}

A quantitative discrete model has been proposed to describe the interaction
of effector cells and cells of a growing tumor. The model adequately
describes the kinetics of tumor growth and regression over a wide interval
of tumor growth rate. Hypothesizing that cytotoxic effector cells are
responsible for the antitumor reactivity, we have been found that the model
can account for many phenomena observed in vivo. According to our model the
dynamics of  the tumor-growth the slow Malthusian growth rate  is the most
predictable. In this case, results of our discrete model, which can be
estimated as a pseudo-Euler's method, is similar to that of continuous-time
models [1,2]. For a malignant and potentially lethal tumor when the time of
number-doubling of tumor cells $T=\ln 2/\alpha \leq \Delta \tau $ the
metastatic disease has been realized. In this case the numbers of effector
and tumor cells fluctuates drastically and the immunity+tumor system is
unpredictable. Besides that, any chaotic system posses a property called
sensitive dependence on initial conditions, and this property precludes
long-term predictions. Nevertheless, attempts to predict the future in
chaotic time series in clinical experiments in vivo can provide useful
information about system generating time series and scenario for
immunotherapy and may eventually be used as an alternative procedure for
identifying some parameters which measurement impossible in vivo. Special
routines, known as nonlinear forecasting programs, may be developed for such
predictions on the basis of the proposed discrete models (2-dimensional
maps).We believe such a system could help in understanding self-organization
properties of immune systems and real temporal series in immunology of tumor
growth. The model may also be applicable to other immunology processes,
where the target of effector cells may be any biological material such as
bacteria, viruses, tumor cells, etc.

\subsection{Literature}

1. V.A. Kuznetsov, I.A. Makalkin, M.A.Taylor, A.S. Perelson Bullet. Math.
Biol. 1994, v.56, N2, p.295-321.

2. H. Mayer, K.S. Zaenker, U. an der Heiden. Chaos, 1995, v.5, N1, p.155-161.

3. S. Michelson, J.T. Leith . Bullet. Math. Biol. 1995, v.57, N2, p.345-366.

\epsfbox{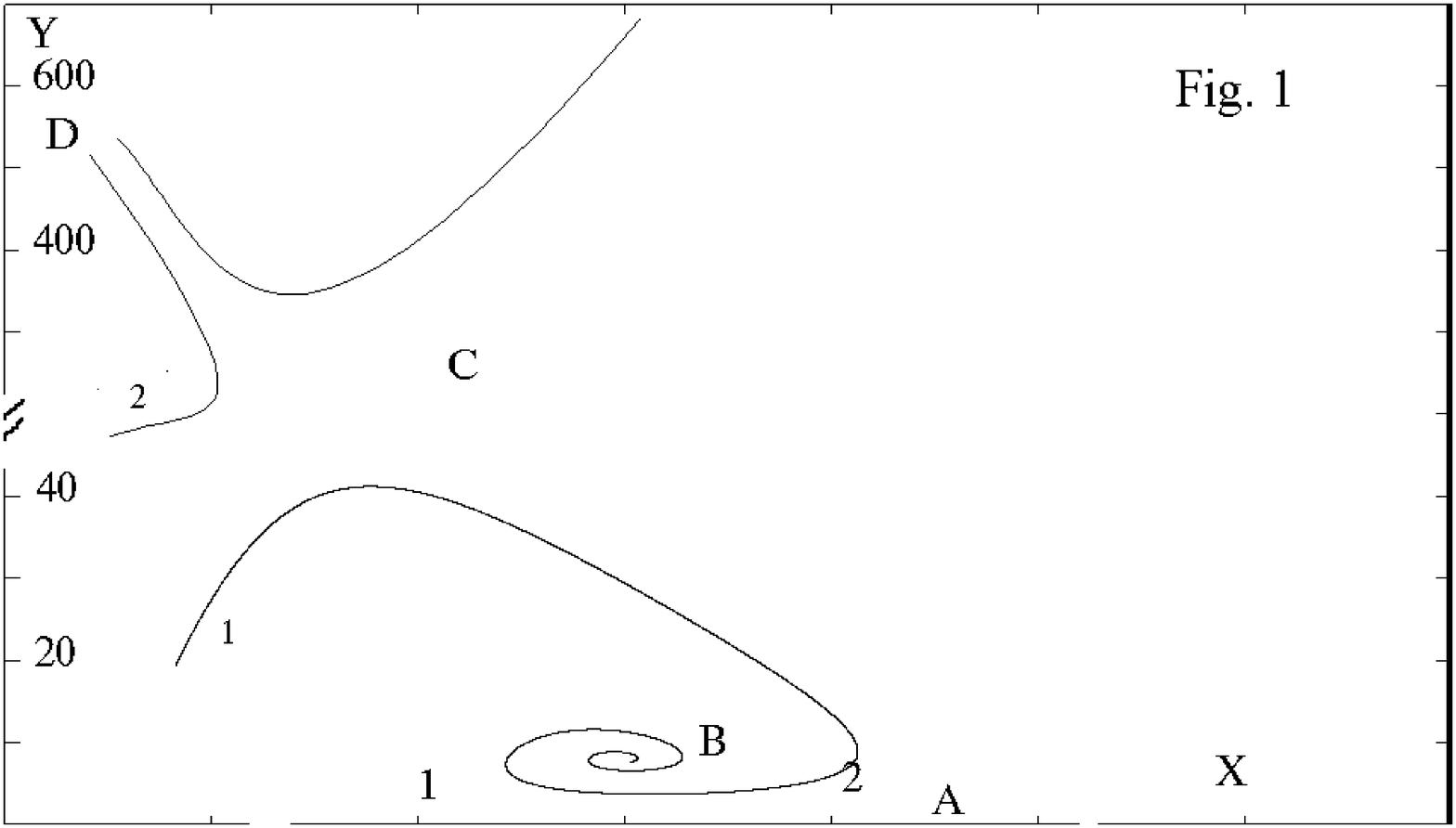}
\epsfbox{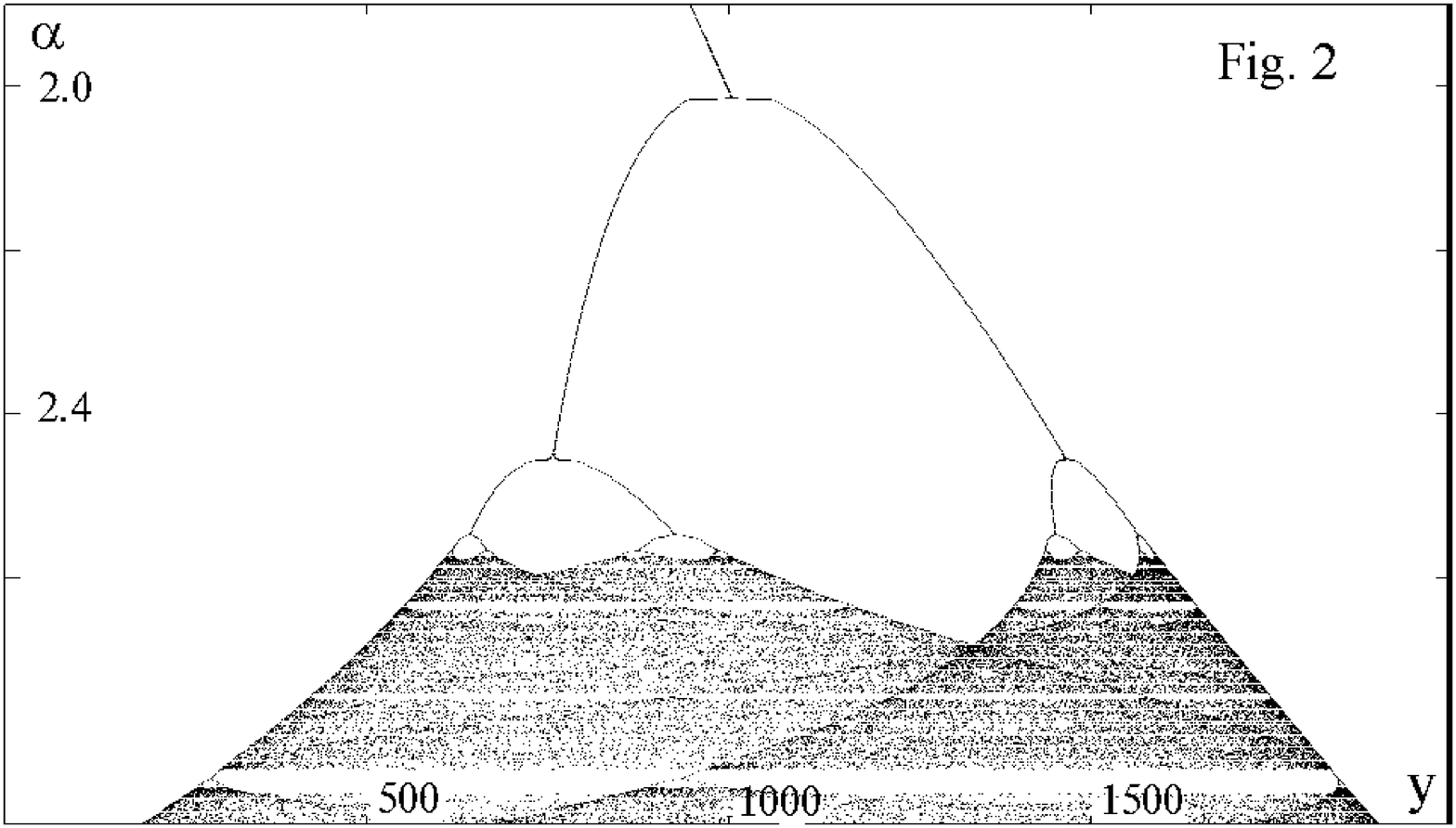}
\epsfbox{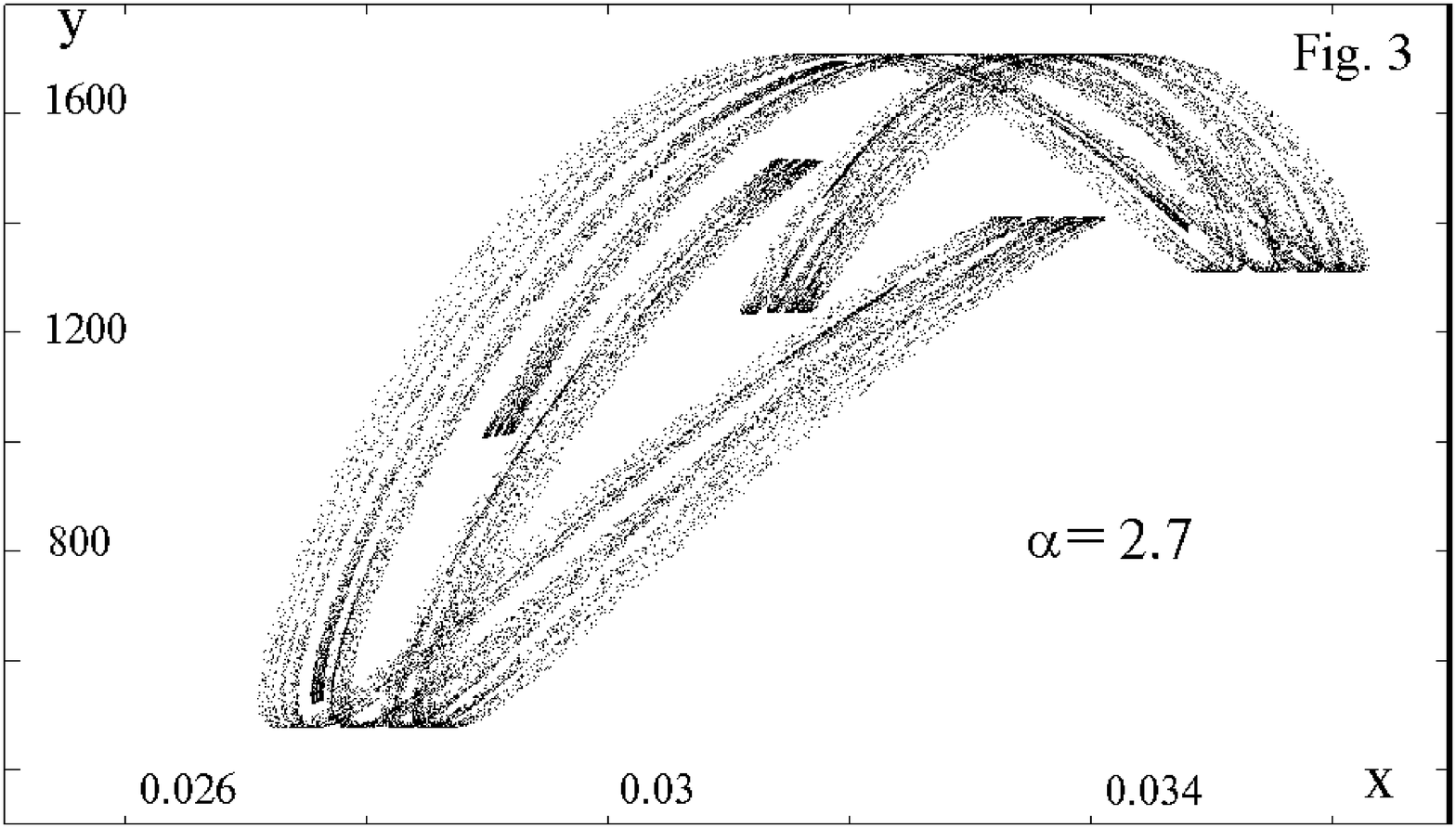}

\end{document}